\documentclass[aps,twocolumn,pre,showkeys,showpacs,eqsecnum]{revtex4-1}
\usepackage{graphpap}
\usepackage[dvips]{graphicx}
\usepackage[dvips]{graphics}
\usepackage{color}
\usepackage[normalem]{ulem} 
\usepackage {ulem} 

\begin{document}

\title{Graphene-based charge sensors}
 \author{C. Neumann$^{1,2}$, C. Volk$^{1,2}$, S. Engels$^{1,2}$, and C. Stampfer$^{1,2}$}
\affiliation{
$^1$JARA-FIT and II. Institute of Physics B, RWTH Aachen University, 52074 Aachen, Germany\\
$^2$Peter Gr\"unberg Institute (PGI-9), Forschungszentrum J\"ulich, 52425 J\"ulich, Germany
}

\date{ \today}

\begin{abstract}
We discuss graphene nanoribbon-based charge sensors and focus on their functionality in the presence of external magnetic fields and high frequency pulses applied to a nearby gate electrode. The charge detectors work well with in-plane magnetic fields of up to 7~T and pulse frequencies of up to 20~MHz. By analyzing the step height in the charge detector's current at individual charging events in a nearby quantum dot, we determine the ideal operation conditions with respect to the applied charge detector bias. Average charge sensitivities of $1.3\times 10^{-3}e/\sqrt{\textrm{Hz}}$ can be achieved. Additionally, we investigate the back action of the charge detector current on the quantum transport through a nearby quantum dot. By setting the charge detector bias from 0 to 4.5~mV, we can increase the Coulomb peak currents measured at the quantum dot by a factor of around 400. Furthermore, we can completely lift the Coulomb blockade in the quantum dot.
\end{abstract}

\maketitle

\newpage
\section{Introduction}
Charge sensors play an important role in low-dimensional electronic circuits, where detecting changes of localized charge states
are crucial and challenging tasks.
In fact, nanoelectronic systems \emph{i.e.} electronic systems with reduced dimensions show a variety of interesting physics including Coulomb blockade~\cite{Kou97}, Kondo effect~\cite{Gol98} or Fano resonances~\cite{Joh04}, all closely related to the localization of electronic charge. Read out and manipulation of isolated electrons are key elements for studying and exploiting these phenomena. Along this line charge detectors based on quantum point contacts (QPCs)~\cite{Wee88} have extensively been used in two dimensional electron systems~\cite{Fie93}. In particular III/V heterostructures have been used as host materials for QPCs. In such devices coherent spin and charge manipulation~\cite{Hay03,Pet05}, full counting statistics~\cite{Gus06}, time resolved charge detection~\cite{Gus06,Fuj06} and controllable coupling to different quantum devices~\cite{Elz04,Shi09,Fre12} have been demonstrated. Moreover, QPC-based charge detectors are regularly used to read out spin qubits realized in double quantum dot systems in GaAs/AlGaAs heterostructures~\cite{Elz03,Blu10,Now11,Shu12}. In these experiments the charge detection fidelity is of great interest in order to maximize the read out speed. The detection fidelity can be optimized by increasing the pinch-off slope and the capacitance between the QPC and the investigated device. While the capacitance can be tuned by reducing the distance between the charge detector and the system of interest (see illustration in Fig. 1(a)), the slope can be increased by replacing the QPC by a single electron transistor (SET). Recently, the use of charge detectors has been extended to hybrid systems where for example a nanowire quantum dot was probed by an underlying QPC detector~\cite{Cho12,Shu08} or a metallic SET was used to detect charging events on a carbon nanotube quantum dot~\cite{Got08}.

\begin{figure}[hbt]\centering
\includegraphics[draft=false,keepaspectratio=true,clip,%
                   width=0.9\linewidth]%
                   {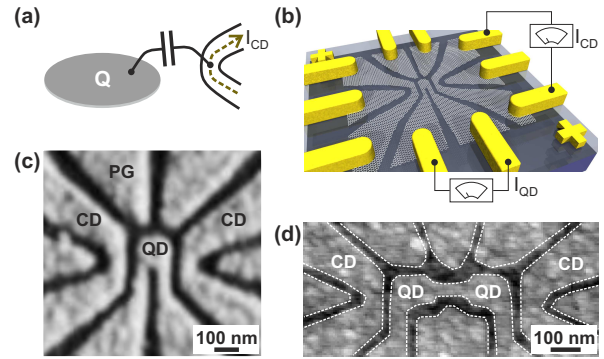}
\caption[FIG1]{(color online) (a) Schematic illustration of a quantum dot (QD) with localized charge Q and capacitively coupled charge detector (CD). (b) Illustration of a contacted graphene QD with a nearby graphene nanoribbon CD. (c) and (d) Scanning force microscope (SFM) image of an etched graphene single (c) and double (d) QD device surrounded by side gate electrodes (See for example the plunger gate (PG) in panel (c).) and CDs.}
\label{fig1}
\end{figure}

More recently, it has also been shown that narrow graphene ribbons can be used as well-working charge detectors~\cite{Gue08}. This approach has  been employed to perform charge sensing on individual graphene quantum dots~\cite{Gue08,Mue12,Wan10,Fri11,Vol13}, including time resolved detection of charging events on such systems~\cite{Gue11}. Additionally, a carbon nanotube-graphene hybrid device has recently been demonstrated where the charge state of a carbon nanotube quantum dot can be detected by the current through a nearby graphene nanoribbon-based charge sensor~\cite{Eng13}.

In particular, graphene attracted increasing interest in the last years, which is mainly due to its remarkable electronic properties such as high carrier mobilities, suppression of direct backscattering and low intrinsic spin noise, which makes graphene an interesting candidate for future electronics and quantum information technology~\cite{Tra07,Los98}. For example, graphene based quantum dots (QD) promise weaker hyperfine coupling as well as weaker spin orbit interaction compared to state-of-the-art III/V heterostructure devices~\cite{Min06, Hue06}. A key challenge when creating graphene based electronic devices is the absence of a band gap in this material and the phenomenon of Klein tunneling, making it difficult to electrostatically confine electrons~\cite{Kat06}. However, structuring graphene on sub-micron scales yields a possibility to overcome this problem as a mainly disorder dominated energy gap opens~\cite{Han07, Sta09, Liu09, Gal10, Han10, Ter11}. Thus, graphene nanoribbons, single-electron-transistors, QDs and double quantum dots have been successfully fabricated and investigated over the past years~\cite{Tod09, Ihn10, Sta08, Pon08, Mos10, Vol11, Gue12}. Furthermore, it has been shown that it is possible to fabricate graphene quantum dots with integrated graphene nanoribbon-based charge detectors in a single fabrication step~\cite{Gue11} leading to a more reliable and reproducible fabrication technology for potentially high quality charge sensors.

In this article, we discuss the fabrication of graphene nanoribbon-based charge detectors (section II) and characterize their behaviour in transport measurements (section III). We show that the charge detector retains its functionality under applying of a magnetic fields and square voltage pulses on a nearby graphene side gate. Finally, we investigate back action effects of the charge detector on a probed graphene QD.

\begin{figure}[t]\centering
\includegraphics[draft=false,keepaspectratio=true,clip,%
                   width=1.0\linewidth]%
                   {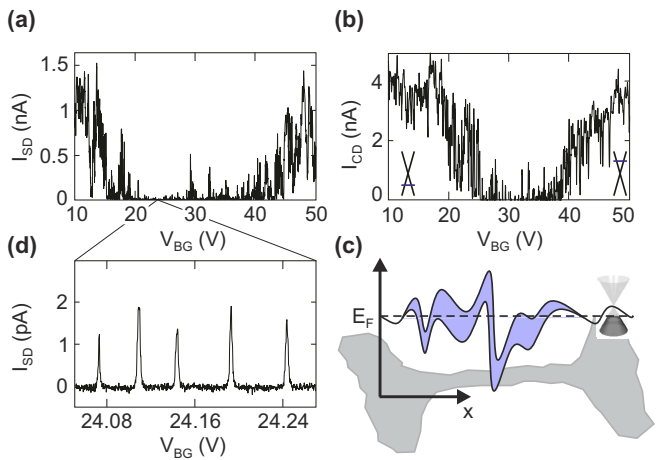}
\caption[FIG2]{(color online) (a) Source-drain current $I_{SD}$ measured at a quantum dot as a function of back gate voltage $V_{BG}$. A transport gap, where the current is strongly suppressed, is observed in the range $V_{BG}=24~V$ to $38~V$. (b) Similar measurement as in
panel (a) but measured on the charge detector. The inset highlights the hole and electron dominated transport regions. (c) Schematic illustration of the disorder induced transport gap in graphene nanoribbons. A confinement induced energy gap (blue area) opens leading to tunneling barriers separating charge puddles arising from disorder induced potential fluctuations. The Fermi level is indicated by the dashed line.
(d) Close up of the measurement shown in panel (a). Inside the transport gap distinct Coulomb resonances are observed.}
\label{fig2}
\end{figure}

\begin{figure}[t]\centering
\includegraphics[draft=false,keepaspectratio=true,clip,%
                   width=0.99\linewidth]%
                   {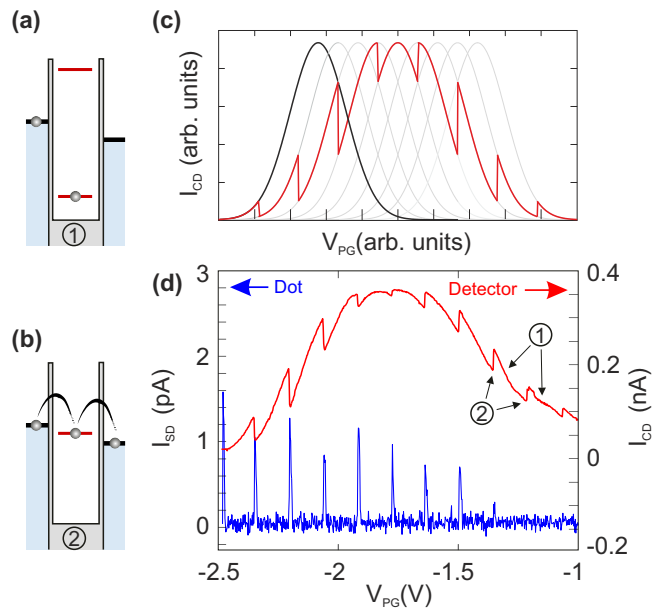}
\caption[FIG3]{(color online) (a) and (b) schematic energy diagram of a QD in (a) the Coulomb blockade regime and (b) regime where transport is possible via sequential tunneling of electrons through the QD. (c) Schematic illustration of the operation principle of the CD. Each individual charging event on the QD shifts the CD resonance (black and gray lines) due to the capacitive coupling of both devices. This mechanism results in steps in the gate dependent measurement (red line). (d) Simultaneous measurement of the QD and CD current as a function of the plunger gate voltage $V_{PG}$. The CD resonance shifts are perfectly aligned with the Coulomb resonances of the QD. Even at very low QD currents (\emph{e.g.} at $V_{PG} = -1.22$~V, $V_{PG} = -1.07$~V) the CD can resolve individual charging events.}
\label{fig3}
\end{figure}

\begin{figure*}[t]\centering
\includegraphics[draft=false,keepaspectratio=true,clip,%
                   width=0.9\linewidth]%
                   {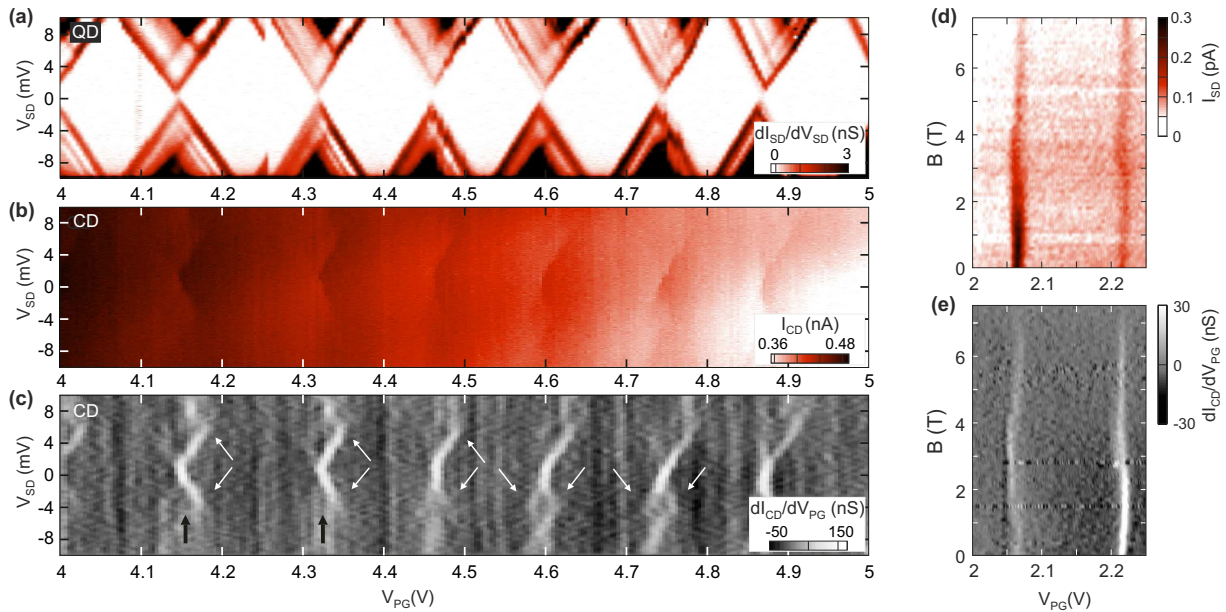}
\caption[FIG4]{(color online) (a) Differential conductance $dI_{SD}$/$dV_{SD}$ as a function of $V_{SD}$ and $V_{PG}$. Coulomb diamonds as well as numerous excited states are visible. (b) and (c) Simultaneously recorded $I_{CD}$ and differential transconductance, $dI_{CD}$/$dV_{PG}$. The charge detector current $I_{CD}$ shows an abrupt step for each charging event in the QD (see panel (b)). Additionally, the charge detection differential transconductance (c) exhibits features which can be associated with excited states (see white arrows in (c)). (d) Magnetic field dependence of two Coulomb resonances at $V_{SD} = -1.5$~mV. (e) Both resonances can clearly be resolved in the CD differential transconductance over the entire shown magnetic field range.}
\label{fig4}
\end{figure*}

\section{Device fabrication}
The device fabrication is based on mechanical exfoliation of natural graphite. The ultra-thin graphite flakes are placed on an insulating 290~nm thick silicon oxide (SiO$_2$) layer on top of highly doped silicon (Si$^{++}$). To identify graphene flakes optical microscopy complemented by Raman spectroscopy is employed~\cite{Fer06, Gra07}. Individual graphene flakes are subsequently nanostructured by electron beam lithography (EBL) and reactive ion etching (RIE) with an Ar/O$_2$ plasma~\cite{Mol07}. The graphene nanostructures are contacted by an additional EBL step followed by metalization and lift-off (see illustration in Fig.~1(b)). Our contacts consist of 5~nm Cr and 50~nm Au. In Figs. 1(c) and 1(d) we show two scanning force microscope (SFM) images of both, (c) a graphene quantum dot (QD), and (d) a double quantum dot device with integrated graphene charge detectors (CDs). In both devices the charge detectors are capacitively coupled to a graphene island hosting localized electrons.

Here, we will mainly focus on graphene CDs coupled to graphene single quantum dot structures as depicted in Figs. 1(b) and 1(c). Our graphene QDs have diameters of around 100~nm and they are connected to source and drain leads by narrow graphene constrictions which act as tunable tunneling barriers~\cite{Sta08}.
Two nanoribbons with a width of about 70~nm, located at either side of the QD, act as charge detectors.
By applying a reference potential to the graphene charge detectors, they can also be used as lateral gates, especially for tuning the transparency of the tunneling barriers.
Additional plunger gates (PG) allow to electrostatically tune the QD potential as well as the tunneling rates.
The highly doped Si substrate acts as a back gate and can be employed to tune the overall Fermi level in the graphene device.

\section{Measurements and discussion}
All transport measurements presented in this article are performed in a dilution refrigerator with a base temperature below 20~mK. Home built amplifiers are used to detect currents with noise levels of around 10~fA/$\sqrt{\textrm{Hz}}$. For the pulse gating experiments a Tektronix AWG 7082C is used to provide rectangular pulses with rise times of about 250-300~ps.

\subsection{Device characterization}
In Figs.~2(a) and 2(b) we show the transport characteristics of a graphene quantum dot and a graphene nanoribbon and quantum dot on a large energy, \emph{i.e.} large back gate voltage range, respectively. Here, the current is measured while varying the back gate voltage from 10 to 50~V. For small back gate voltages both devices show hole-dominated transport, while for large back gate voltages the transport is electron-dominated (see also insets in Fig.~2(b)). These two regions are separated by the so-called transport gap, where the measured current is strongly suppressed. The transport gap is located at positive gate voltages (around 20 to 37~V in Fig. 2(a) and 27 to 37~V in Fig. 2(b)) indicating a significant p-doping of both structures, which is commonly observed in etched graphene nanostructures~\cite{Ter11,Gue12}. This p-doping arises most likely due to polymer resist residues and/or oxygen atoms bound to the graphene edges coming from the Ar/O$_2$ plasma etching process~\cite{Ryu10}.
In the transport gap regime the electronic transport is dominated by stochastic Coulomb blockade where lateral confinement in combination with a significant disorder potential, arising from both (i) bulk disorder and (ii) edge roughness, play an important role (see illustration in Fig.~2(c))~\cite{Sta09,Mol10}. The observed large-scale current fluctuations inside the gap region originate from local resonances in the graphene constriction. In Fig.~2(d) we show a close up of the back gate characteristics of the graphene QD inside the transport gap, highlighting individual Coulomb resonances. Transport is blocked if no QD state is aligned inside the bias window (Coulomb blockade, see Fig.~3(a)). Individual Coulomb resonances are observed if a state is aligned between the chemical potentials of the source and drain leads (Fig. 3(b)).

Individual conductance resonances in the transport characteristics of a graphene nanoribbon can be used to detect charging events on a capacitively coupled QD close by. Whenever the overall charge of the QD changes, the electrostatic potential in the nanoribbon-based charge detector is shifted which results in a conductance step of the CD (Fig.~3(c)). Thus, individual charging events in the QD can be probed by measuring the current passing through the nanoribbon, which consequently acts as a charge detector~\cite{Gue08,Wan10,Fri11}. In Fig.~3(d) the simultaneously measured current through the quantum dot ($I_{SD}$) and the charge detector ($I_{CD}$) are shown as a function of the plunger gate voltage $V_{PG}$. The steps in $I_{CD}$ are well aligned with the Coulomb peaks in $I_{SD}$. The CD can even resolve charging events that cannot be measured by the direct current since the Coulomb peaks vanish in the noise of the $I_{SD}$ signal (see $V_{PG}>-1.5~$V in Fig. 3(d)).

In Fig. 4 (a) we show so-called Coulomb diamonds in the differential conductance measured on the QD. Signatures of transport through excited states in the QD are observed as faint lines running parallel to the diamond edges. The CD signal in Figs.~4(b) and 4(c) shows features well aligned with the Coulomb resonances visible in Fig.~4(a). Interestingly, for the two left Coulomb diamonds (marked by the vertical black arrows in Fig. 4(c)) the dominant tunneling barrier changes when going from negative to positive $V_{SD}$, indicating a rather strong capacitive coupling of the QD tunnel barriers to the source and the drain lead. Apart from the ground state transport also transport via excited state transitions can be identified in the differential transconductance of the CD, highlighted by white arrows.

\begin{figure}[t]\centering
\includegraphics[draft=false,keepaspectratio=true,clip,%
                   width=1.0\linewidth]%
                   {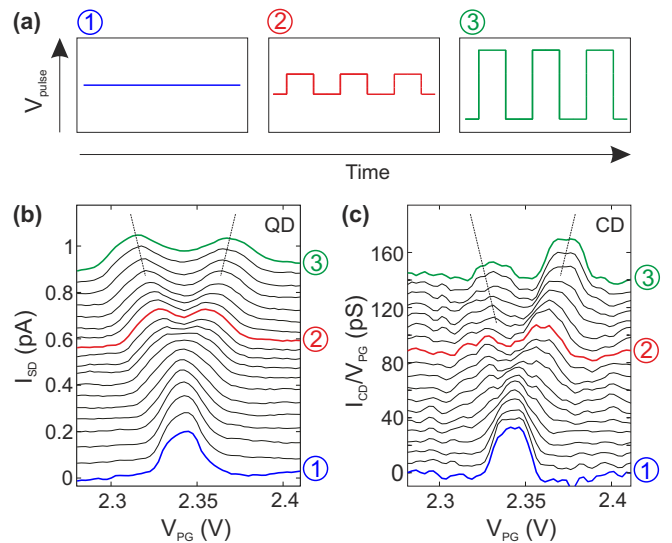}
\caption[FIG5]{(color online) (a) Schematics of rectangular pulse sequences with a duty cycle of 50\% and increasing amplitude. (b) Coulomb resonance under the influence of a 20 MHz pulse sequence applied to the plunger gate. The resonance splits linearly into two peaks with increasing pulse amplitude (see dashed lines). (c) Corresponding CD differential transconductance, $dI_{CD}$/$dV_{PG}$, reflecting consistently
the peak splitting.}
\label{fig5}
\end{figure}

\subsection{Charge detection at finite magnetic fields and pulsed-gates}

Next we demonstrate that the graphene nanoribbon-based CD also works over a large range of external magnetic fields.
This is important for detecting individual spin states in graphene QDs. Please note that it has been shown that magnetic fields may strongly alter the transport properties of graphene nanoribbons~\cite{Gue11a}, rendering this a non-trivial task.
In Fig. 4(d) two Coulomb resonances are measured for an in-plane magnetic field range of 0 to 7.4~T. Both resonances are also present in the transconductance in Fig.~4(e) with similar resolution.
So far, all results have been obtained from DC measurements. In the measurements shown in Fig. 5, a pulsed-gate technique was used to manipulate the transport through the QD on nanosecond time scales~\cite{Fuj01,Dro12,Vol13}. In these measurements a rectangular pulse applied to the plunger gate (see "PG" in Fig. 1(c)) located in the vicinity of the QD. Importantly, the voltage pulse shifts the QD potential by the electrostatic coupling. In Fig. 5(b) the evolution of a Coulomb resonance for increasing pulse amplitude at a constant frequency of 20~MHz is plotted. The Coulomb resonance splits into two peaks with increasing pulse amplitude. The equivalent height of the two peaks after splitting is in good agreement with the pulse duty cycle of 50\% (see Fig. 5(b)). This splitting is also observed in the charge detector transconductance as shown in Fig.~5(c).

\begin{figure*}[t]\centering
\includegraphics[draft=false,keepaspectratio=true,clip,%
                   width=0.9\linewidth]%
                   {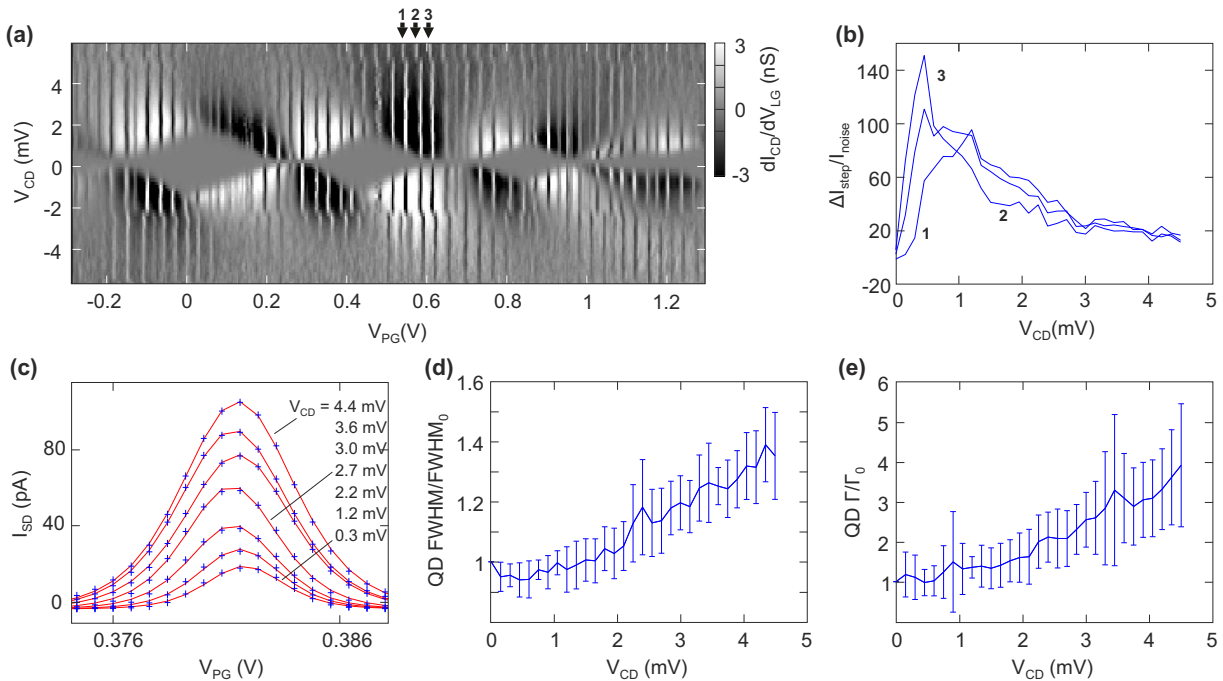}
\caption[FIG6]{(color online) (a) Differential transconductance of the CD as a function of $V_{CD}$ and $V_{PG}$. The charging events of the QD detected by the CD appear as white and dark vertical lines. Additionally, the CD shows clear Coulomb diamonds arising from the disorder induced isolated islands in the graphene nanoribbon. (b) Charge detector current step height divided by the average noise level of current as function of of $V_{CD}$ of three different charging events in the QD marked by the three black arrows in panel (a). The highest steps occur at the onset of the current at the edges of the Coulomb diamonds in the CD. (c) Evolution of a Coulomb resonance of the QD with increasing $V_{CD}$. Red lines show the fit to the experimental data (blue crosses) using the expression $I_{SD}(V_{PG}) = I_{max}/{cosh^2[(V_{PG}-V_{res})/a]}$. (d) and (e) FWHM (d) and peak current (e) averaged over 8 Coulomb resonances as a function of $V_{CD}$ (normalized to $V_{CD} = 0 $~mV). The FWHM increases by 50\% at $V_{CD} = 4.5$~mV while an average increase of the peak current by 400\% can be observed.}
\label{fig6}
\end{figure*}

\subsection{Back action}
In the measurements discussed in the previous sections the charge detection bias ($V_{CD}$) was set such that charging events in the QD could be easily detected. In the following we will investigate (i) the influence of the CD bias voltage on the detection sensitivity as well as (ii) the back action of the detection on the transport through the quantum dot.
In Fig.~6(a) we show the transconductance of the detector measured as function of detector bias ($V_{CD}$) and plunger gate voltage. Remarkably,
for small bias voltages we observe well-resolved Coulomb diamonds with charging energies on the order of 1.5~meV. This is in perfect agreement with the nature of the resonances in the charge detector nanoribbon. Inside the diamonds, where transport through the nanoribbon is (Coulomb) blocked the CD is completely insensitive to charging events in the nearby QD.

In order to detect QD states over an extended $V_{PG}$ range the CD must be operated at a bias value outside the Coulomb blockade regime. However, increasing $V_{CD}$ leads to an overall increased current through the CD. This broadens the CD resonances, which results in a decreasing signal-to-noise ratio for detecting individual charging events.
In Fig.~6(b) we show the signal-to-noise ratio of three conductance steps of the CD investigated for a $V_{CD}$ range of 0 to 5~mV. The signal-to-noise ratio is best for low $V_{CD}$ as long as the CD is not blocked due to Coulomb blockade. From Fig. 6(b) we extract an average charge sensitivity of $1.3 \times 10^{-3} e/\sqrt{\textrm{Hz}}$ at $V_{CD}$ = 0.5 mV. Please note that the charge sensitivity strongly scales with $V_{CD}$ and the slope of the CD resonance where the charging event in the QD occurs.

Furthermore,  the transport through the QD is strongly affected by the charge detector current. The Coulomb resonances of the QD broaden with increasing charge detector bias which is due to back action effects. We phenomenologically investigate this behavior by fitting the Coulomb peak in the QD by $I_{SD}(V_{PG}) = \frac{I_{max}}{cosh^2[(V_{PG}-V_{res})/a]}$, with the fit parameters $V_{res}$, $I_{max}$ and $a$ to different Coulomb resonances located in the regime of $V_{PG}$ = 0.3 to 0.5~V. An exemplary evolution of a Coulomb resonance with increasing $V_{CD}$ is displayed in Fig. 6(c). Fig. 6(d) shows the FWHM averaged over 8 Coulomb resonances depending on $V_{CD}$. Varying $V_{CD}$ from 0 to 4.5~mV increases the FWHM by around 50\%. At the same time the average peak height rises by about 400\% (see Fig. 6(e)). The strong back action of the CD onto the QD can also nicely be seen in Fig. 7, where the low-bias current through the QD is in log-scale plotted as function of $V_{PG}$ and $V_{CD}$. With increasing $V_{CD}$ the Coulomb resonances in the QD broaden until the Coulomb blockade is completely lifted.

This dependency on the current flowing through the CD indicates that the increase and broadening of the Coulomb resonances originate from
noise and fluctuations in the CD nanoribbon~\cite{Gue11}. Coupling of the nanoribbon to the QD via phonons seems less plausible as the phonons would have to couple via the SiO$_2$ substrate due to the destroyed graphene lattice. Consequently, it is likely that photons play an important role and that processes related to photon assisted tunneling are responsible for lifting the Coulomb blockade in the transport through
the graphene QD.
In summary, the best conditions for charge sensing, \emph{i.e.} for detecting individual charging events in the nearby QD found at $V_{CD}$ slightly above the Coulomb blockade regime of the CD. In this case all QD resonances can be probed with a good signal-to-noise ratio, while the back action on the QD is kept small.

\begin{figure}[t]\centering
\includegraphics[draft=false,keepaspectratio=true,clip,%
                   width=0.9\linewidth]%
                   {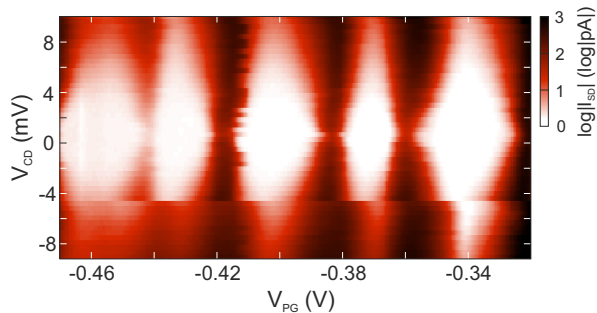}
\caption[FIG7]{(color online) Current through the quantum dot ($I_{SD}$) as function of $V_{PG}$ and charge detector bias, $V_{CD}$ for constant $V_{SD} = 0.5$~mV. The Coulomb resonances of the Quantum dot broaden with increasing $V_{CD}$ such that a diamond-shaped pattern can be observed. The Coulomb blockade is completely lifted for $V_{CD}$ exceeding 9 mV.}
\label{fig7}
\end{figure}

\section{Conclusion}
We have characterized graphene nanoribbon-based charge detectors. By bias spectroscopy measurements we have shown that excited states of a nearby quantum dot are resolved with the graphene charge detector. Furthermore, the charge detector was successfully tested in experimental setups where in plane magnetic fields up to more than 7~T and pulsed-gates at 20~MHz were used to manipulate individual QD states. We reach average charge sensitivities on the order of 1.3$\times 10^{-3} e/\sqrt{\textrm{Hz}}$. The charge detector can be operated in regimes with high sensitivity, while the back action onto the investigated QD is kept at a minimum. However, at higher charge detector bias voltages, we observe complete lifting of the Coulomb blockade regime in the probed quantum dot.
A detailed understanding of the graphene-based charge sensors may open new fields of applications, in particular in view of the truly two-dimensional nature of these sensors.

\section*{Acknowledgement}
We acknowledge S. Trellenkamp for support with electron beam lithography and U. Wichmann for help with the low-noise measurement electronics. We thank F. Haupt, B. Terr\'{e}s, and A. Epping for helpful discussions. Support by the DFG (SPP-1459 and FOR-912), JARA seed fund, and the ERC are gratefully acknowledged.

\newpage

\newpage

\newpage

\newpage

\newpage

\newpage

\newpage

\clearpage

\end{document}